%% file: main.tex
\documentclass[10pt,conference,letterpaper]{IEEEtran}
\IEEEoverridecommandlockouts
\IEEEpubid{\makebox[\columnwidth]{ 979-8-3503-7608-1/24\$31.00 \copyright2024 IEEE \hfill} \hspace{\columnsep}\makebox[\columnwidth]{ }}

\input{sec/00preamble}

\input{sec/00preamblelistingsconf}

\begin{document}

\title{LLM-aided explanations of EDA synthesis errors}

\IEEEaftertitletext{\vspace{-2.3\baselineskip}}

\author{\IEEEauthorblockN{Siyu Qiu}
\IEEEauthorblockA{\textit{University of New South Wales}\\
Sydney, NSW Australia \\
siyu.qiu1@student.unsw.edu.au}
\and
\IEEEauthorblockN{Benjamin Tan}
\IEEEauthorblockA{\textit{University of Calgary}\\
Calgary, AB Canada \\
benjamin.tan1@ucalgary.ca}
\and
\IEEEauthorblockN{Hammond Pearce}
\IEEEauthorblockA{\textit{University of New South Wales} \\
Sydney, NSW Australia \\
hammond.pearce@unsw.edu.au}
\thanks{The research in this work was supported in part by Intel Corporation and in part by Woodpecker Technologies. 
This work does not in any way constitute an Intel endorsement of a product or supplier.
}
}

\maketitle

\IEEEpubidadjcol

\begin{abstract}

Training new engineers in digital design is a challenge, particularly when it comes to teaching the complex electronic design automation (EDA) tooling used in this domain. 
Learners will typically deploy designs in the Verilog and VHDL hardware description languages to Field Programmable Gate Arrays (FPGAs) from Altera (Intel) and Xilinx (AMD) via proprietary closed-source toolchains (Quartus Prime and Vivado, respectively). 
These tools are complex and difficult to use---yet, as they are the tools used in industry, they are an essential first step in this space.
In this work, we examine how recent advances in artificial intelligence may be leveraged to address aspects of this challenge.
Specifically, we investigate if Large Language Models (LLMs), which have demonstrated text comprehension and question-answering capabilities, can be used to generate novice-friendly explanations of compile-time synthesis error messages from Quartus Prime and Vivado. 
To perform this study we generate 936 error message explanations using three OpenAI LLMs over 21 different buggy code samples. These are then graded for relevance and correctness, and we find that in approximately 71\% of cases the LLMs give correct \& complete explanations suitable for novice learners.
\end{abstract}


\input{sec/01introduction}

\input{sec/02literature}

\input{sec/03method}

\input{sec/04results}

\input{sec/05conclusions}

\clearpage

\bibliographystyle{IEEEtran}
\bibliography{ref/benhamram}

\section*{Appendix}

Figure~\ref{fig:prompts} shows the system and user prompt templates used to generate the error explanations in this study. Prompting was the same for all 3 OpenAI models.

\begin{figure}[h]
    \centering
    \begin{lstlisting}
You are a helpful assistant which debugs RTL and HDL code in Verilog and VHDL. Do not provide code in your answer. Explain what has gone wrong and why a bug is occurring, but do not attempt to fix the bug yourself.
    \end{lstlisting}
    \vspace{-2mm}
    {\footnotesize (a) System prompt.}
    \begin{lstlisting}
Error message: %s

Full code file:```%s```

What is the bug and why is it occurring?
    \end{lstlisting}
    \vspace{-2mm}
    {\footnotesize (b) Error \& Code (E\&C) prompt template.}
\begin{lstlisting}
Error message: %s

Error line:```%s```

Full code file:```%s```

What is the bug and why is it occurring?
    \end{lstlisting}
    \vspace{-2mm}
    {\footnotesize (c) Error, Code, \& Line (EC\&L) prompt template.}
    \caption{LLM prompts}
    \label{fig:prompts}
\end{figure}

Figure~\ref{fig:explanation:too-much-help} shows an explanation with `solution is provided'. 

\begin{figure}[h]
    \centering
    \begin{lstlisting}
The error message is indicating that there's syntax error near the 'elsif' keyword. 

By looking at your VHDL code, the problem is that you are missing a semicolon at the end of the line after the assignment operation when the rst is '1'. In VHDL, every statement ends with a semicolon. 

The line of the code:
```
if rst = '1' then
    data_out <= (others => '0')     --Missing semicolon here
```

Should be corrected with a semicolon at the end as:
```
if rst = '1' then
    data_out <= (others => '0');    --Add semicolon here
``` 

Your synthesizer (the tool that's processing your VHDL code) is seeing the 'elsif' as being part of the previous line (because there was no semicolon to end it) and is confused as this doesn't make sense and therefore throws an error. This is why the error message is suggesting there may be a problem near the 'elsif'. 

After adding the semicolon at the end of the line in question, the issue should be resolved.
    \end{lstlisting}
    \vspace{-2mm}
    {\footnotesize Scores---Concept accurate: \textbf{Yes}; Any inaccuracies: \textbf{No}; Relevant: \textbf{Yes}; Correct \& complete: \textbf{Yes}, Solution is provided: \textbf{Yes}}
    \vspace{-1mm}
    \caption{A `good' error explanation for Bug 1 (Figure~\ref{fig:bad-error}) generated by \texttt{gpt-4}. However, this explanation is flagged `Solution is provided' as it contains a direct copy/paste solution on Lines 12-15---i.e., the explanation helps `too much' according to constructivism in pedagogy.}
    \label{fig:explanation:too-much-help}
\end{figure}

Figure~\ref{fig:explanation:relevant-vs-concept-accurate} illustrates the difference between an explanation flagged as `relevant' versus one flagged as `concept accurate', i.e. the two metrics do not necessarily imply each other.

\begin{figure}[h]
    \centering
    \begin{lstlisting}
The error message "syntax error near elsif" is indicating that there's a syntax issue near the "elsif" keyword in your code. 

In this case, the error is triggered by the misuse of the "rising_edge" function in VHDL. The function expects a signal as its argument and checks whether there's a rising edge (low to high transition) on that signal. It's typically used with a clock signal inside a process. 

However, the issue here is that the "elsif" keyword is being used in conjunction with the "rising_edge" function. In VHDL, you don't typically use "elsif" with "rising_edge". Instead, the usual construct would be "if rising_edge(clk)" directly following the process declaration. 

It's likely that the "elsif" is intended to make the data_out assignment happen based on a rising edge event on the clock, but after a reset. The correct usage would be to nest "if rising_edge(clk)" inside the "if rst = '1'" condition, as the first to be processed.

Please remember that you should place the reset and clock conditions under separate if conditions to avoid ambiguities in the descriptions.
    \end{lstlisting}
    \vspace{-1mm}
    {\footnotesize (a) `Bad' explanation by \texttt{gpt-4} which is relevant but with inaccurate concepts, linking to issues with clock edges instead of the actual syntax fault\\Scores---Concept accurate: \textbf{No}; Any inaccuracies: \textbf{Yes}; Relevant: \textbf{Yes}; Correct \& complete: \textbf{No}, Solution is provided: \textbf{No}}

    \begin{lstlisting}
The bug in the code is that the `elsif` statement is missing a semicolon at the end of the assignment in the process. This is causing a syntax error because the VHDL syntax expects a semicolon after each statement within the `if-elsif` structure.
    \end{lstlisting}
    \vspace{-1mm}
    {\footnotesize (b) `Bad' explanation by \texttt{gpt-3.5-turbo} which has the right concepts but with irrelevant info. (the elsif has a semicolon on its later clause).\\Scores---Concept accurate: \textbf{Yes}; Any inaccuracies: \textbf{Yes}; Relevant: \textbf{No}; Correct \& complete: \textbf{No}, Solution is provided: \textbf{No}}
    \vspace{-1mm}
    \caption{Demonstrating the difference between `relevance' and `concept accurate' with two `bad' error explanations for Bug 1 (Figure~\ref{fig:bad-error})}
    \label{fig:explanation:relevant-vs-concept-accurate}
\end{figure}

\end{document}

%% file: sec/00preamble.tex
\usepackage{cite}
\usepackage{amsmath,amssymb,amsfonts}
\usepackage{algorithmic}
\usepackage{graphicx}
\usepackage{textcomp}
\usepackage{xcolor}
\usepackage{tikz}
\usetikzlibrary{tikzmark}
\usetikzlibrary{shapes,arrows,positioning}
\usepackage{listings}
\usepackage{multirow}
\usepackage{pifont}

\usepackage{soul}

\usepackage[hidelinks]{hyperref}
\usepackage{cleveref}
\usepackage{acronym}
\usepackage{enumitem}
\usepackage{xurl} %

\acrodef{LLM}[LLM]{large language model}
\acrodef{HDL}[HDL]{hardware description language}
\acrodef{EDA}[EDA]{electronic design automation}

\makeatletter
\let\MYcaption\@makecaption
\makeatother
\usepackage[font=footnotesize]{subcaption}
\makeatletter
\let\@makecaption\MYcaption
\makeatother

\def\BibTeX{{\rm B\kern-.05em{\sc i\kern-.025em b}\kern-.08em
    T\kern-.1667em\lower.7ex\hbox{E}\kern-.125emX}}

%% file: sec/01introduction.tex
\section{Introduction}

With increasing demand for digital devices, there is a need for more digital design practitioners. 
However, existing \ac{EDA} tools have a considerably steep learning curve. 
For example, in the FPGA design space, Altera and AMD Xilinx tools are frequently used in educational settings. 
These tool suites are renowned for their difficulty and complexity, particularly for new users. 
Indeed, the combination of new languages, design paradigms, software tools, and hardware requirements can leave novices feeling well and truly ``stumped''~\cite{edwards_experiences_2005}, particularly when the software provides unhelpful messages upon reaching erroneous code (e.g., Figure~\ref{fig:bad-error}). Likewise, educators themselves can struggle with the broad knowledge base required~\cite{pasricha_embedded_2022}.
This is a serious challenge, especially considering the worldwide shortfall in qualified chip designers (in the US alone, estimates have a 67,000 employee shortfall by 2030~\cite{cherney_us_2023}).

\begin{figure}
    \centering
    \begin{lstlisting}[language=VHDL,firstnumber=41]
architecture Behavioral of top1 is 
begin 
    process (clk, rst) begin
        if rst = '1' then
            data_out <= (others => '0')
        elsif rising_edge(clk) then
            data_out <= data_in;
        end if;
    end process;
end Behavioral;
    \end{lstlisting}
    \vspace{-2mm}
    {\footnotesize (a) Snippet of buggy VHDL code}
    \begin{lstlisting}
ERROR: [Synth 8-2715] syntax error near elsif [path/to/bug_1/rtl/top1.vhd:46]
    \end{lstlisting}
    \vspace{-2mm}
    {\footnotesize (b) Vivado's corresponding error message}
    \vspace{-2mm}
    \caption{Example unhelpful error message. It does not describe the real problem (a missing semicolon), and it links to line 46, not the fault on line 45!}
    \label{fig:bad-error}
    \vspace{-4mm}
\end{figure}

It is therefore desirable to see if recent advancements in artificial intelligence may be able to \textit{assist} novice digital hardware designers and accelerate their training, be it in the classroom or perhaps as part of onboarding/professional development. %
In particular, \acp{LLM} have demonstrated considerable capabilities for comprehending text and program code, enabling code generation and explanation. 
Given that one of the most common difficulties noted when learning to program comes from understanding and overcoming compiler error messages~\cite{becker_effective_2016,karvelas_effects_2020} we pose the question: \textbf{Can \acp{LLM} be leveraged to explain error messages from EDA tools?} 

In this work, we thus undertake a proof-of-concept examination, tasking a series of OpenAI LLMs with generating explanations for a series of synthesis-time (i.e. compile-time) bugs commonly encountered by novice digital designers. Our synthetic dataset contains error messages from both Intel's Altera Quartus Prime and AMD's Xilinx Vivado with both VHDL- and Verilog-based designs. 

Crucially, as we aim to up-skill tool users, we desire pedagogically-useful responses (i.e., not automated program repair). 
The \ac{LLM} should assist the user but should not outright solve the issue---per the constructivism pedagogy in computer science~\cite{ben-ari_constructivism_2001}, learners should ``build'' knowledge rather than be simply told answers outright. 
Moreover, the insights from our study can also lay the foundation for other \ac{LLM}-based augmentation of \ac{EDA} tool feedback to improve their readability/actionability and, thus, designer productivity. 
Our contributions include the following:
\begin{itemize}[leftmargin=*]
    \item A new open-source dataset of 21 representative synthesis-time bugs and error messages based on the authors' experiences with teaching introductory digital hardware design.
    \item Using these bugs, results from the first pedagogically focused evaluation of 936 \ac{LLM}-generated bug explanations, finding that $\approx$71\% have `good' explanations.
\end{itemize}

\textbf{Open-source:} All synthetic bugs and generated data is provided at \underline{\url{https://zenodo.org/doi/10.5281/zenodo.10937409}}.

%% file: sec/02literature.tex
\section{Background and Related Work}
\label{sec:background}

\subsection{Large Language Models for hardware design}

\acp{LLM}, trained over large quantities of text scraped from the Internet (including millions of open-source repositories), have demonstrated considerable cross-domain expertise in lexical tasks. While early models such as OpenAI's Codex acted primarily as a kind of ``smart autocomplete,'' more recent training methodologies such as Reinforcement Learning with Human Feedback (RLHF) can create models more capable of following user intent~\cite{ouyang_training_2022}---meaning such models may actually ``follow instructions.''
Currently, leading commercial \acp{LLM} in this space come from OpenAI's ChatGPT family~\cite{openai_chatgpt_2022}, %
which can be ``prompted'' to translate and debug code and provide code explanations using natural language. 

Multiple works have acknowledged the potential for \acp{LLM} to work with hardware design tasks, including for authoring \ac{HDL} code~\cite{pearce_dave_2020,thakur_benchmarking_2023,thakur_verigen_2024,liu_invited_2023}, bug fixing~\cite{ahmad_hardware_2024}, SystemVerilog Assertion generation~\cite{kande_security_2024}, and scripting hardware tool suites~\cite{liu_chipnemo_2023} among others. Some works have even explored how conversational \acp{LLM} like OpenAI's ChatGPT~\cite{openai_chatgpt_2022} can author whole processor designs~\cite{blocklove_chip-chat_2023}.

\subsection{\acp{LLM} for training and education}

Proponents of the technology argue that LLMs, when carefully utilized, can unlock new pedagogical tools and strategies. Kasneci et al. provide a comprehensive survey in this area~\cite{kasneci_chatgpt_2023}, finding that, for example, ChatGPT is already being used for educational methods such as generating tests, quizzes, and flashcards~\cite{dijkstra_reading_2022,gabajiwala_quiz_2022}. %

Given their recency, investigations on the challenges and opportunities provided by \acp{LLM} in the education domain are ongoing~\cite{jalil_chatgpt_2023}. 
Particular attention has been provided on their potential impact in ``CS1'' introductory courses (e.g. \cite{becker_programming_2023,denny_conversing_2023}) and on quality of code explanations for novices~\cite{macneil_automatically_2022,macneil_experiences_2023,macneil_generating_2022}. Of particular interest to us is the potential for \acp{LLM} to aid in error message explanation, where \acp{LLM} are used to help learners overcome compile and runtime errors. Taylor et al.~\cite{taylor_dcc_2024} explored this application for C, where they examined over 64,000 uses of the ChatGPT-enabled C compiler DCC by over 2,500 students. They found that over 90\,\% of compile-time and 75\,\% of run-time error messages had valid explanations.

This motivates our investigation, which explores a similar use case for hardware synthesis rather than software compilation. The DCC compiler provided considerable additional context to the ChatGPT 3.5 LLM in the form of stack traces and templated error message assistance~\cite{taylor_dcc_2024}. Can we find a similar level of assistance but for hardware, simply using the context available in the typical hardware EDA tool suite?

%% file: sec/03method.tex
\section{Digital Design Assistance via LLM prompts}

\begin{figure}[t]
    \centering
    \includegraphics[width=0.7\linewidth]{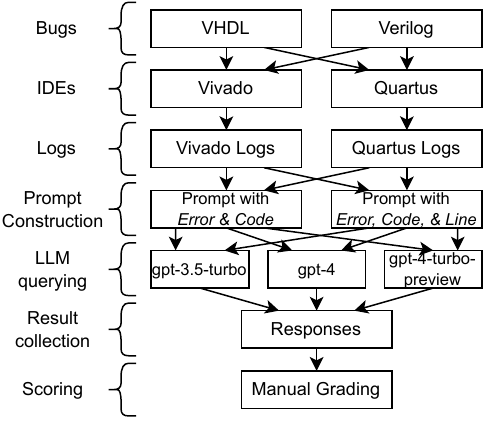}
    \vspace{-2mm}
    \caption{Overall experimentation methodology}
    \vspace{-5mm}
    \label{fig:methodology}
\end{figure}

\subsection{Overview}

This study explores the effectiveness of using LLMs to improve error feedback in Vivado and Quartus.
We use the methodology outlined in Figure~\ref{fig:methodology}.
Firstly, we create a corpus of representative bugs. %
We use these to collect synthesis error messages, %
then use two different prompts to request explanations from selected LLMs. %
Finally, we score the responses. %

\subsection{Defining a corpus of bugs}
\label{sec:method-bugs}
Table~\ref{tab:bugs}'s first four columns present the range of bugs used in this study (the latter four columns present results; see Section~\ref{sec:results}). The bugs were based in general on 10 distinct error categories (e.g., syntax errors, multiple driver errors, type errors and others) that the authors have frequently observed in code written by learners and novice hardware designers. Two of the VHDL bugs did not have an equivalent Verilog representation. The buggy files are short (usually less than 50 lines of comments and code)---e.g.,  Figure~\ref{fig:bad-error} depicts Bug 1.
\begin{table*}
    \centering
    \caption{List of HDL bugs (and their languages) examined in the study, with LLM evaluation results. Each bug had 24 responses in each prompting strategy, and these were manually graded.}
    \vspace{-3mm}
    \label{tab:bugs}
    \resizebox{0.95\linewidth}{!}{\input{fig/error-list}}
    \vspace{-3mm}
\end{table*}

In this work, we focus exclusively on synthesis-time bugs rather than the more complex run-time issues that could occur. The reasons for this are twofold: (1) both the Vivado and Quartus IDEs have limited capabilities in detecting run-time issues, given that they primarily rely on user-provided test-benches in simulation for this purpose. (2) Run-time issues in the novice-focused area will primarily be logic-based, which may be identified by reading simulation waveforms. A user having a simulation error thus has more information available to them than one who is stuck with an inscrutable and unchanging synthesis error message. In future, we plan to extend our study to investigate how \acp{LLM} can be leveraged for more complex debugging and training. 

\subsection{Harvesting Vivado and Quartus error log files}
\label{sec:method-logs}
Quartus and Vivado can synthesize VHDL and Verilog files into bitstreams for FPGAs. When an error occurs, synthesis stops, and the error message is saved to a file---Quartus stores synthesis logs in a file like `\texttt{\small \path{path/to/project/output_files/project.map.rpt}}', and Vivado `\texttt{\small \path{path/to/project/project_1.runs/synth_1/runme.log}}'. 

As these logs also include other information about the tool flow, we extract error messages using regular expressions, scanning for lines beginning with ``Error:'' for Quartus and ``ERROR:'' for Vivado. Further regular expressions can extract details about the error, such as the message, faulty file, and reported error line number.

\subsection{Prompting LLMs for error explanations}
\label{sec:method-llms}
LLMs function by providing an output ``response'' to an input ``prompt'. For this study, we selected three of OpenAI's LLMs, \texttt{\small gpt-3.5-turbo}, \texttt{\small gpt-4}, and \texttt{\small gpt-4-turbo-preview}. These first take a ``System'' prompt to provide overall model guidance, followed by ``User'' prompts which can contain data. As the models evolve over time, we note our usage was on March 29, 2024. 

Recall that the goal of this work is for error explanations not bug repair.
We therefore base our LLM prompting strategy on that used in prior work~\cite{taylor_dcc_2024}, which aimed for pedagogically-focused error message explanations. 
As depicted in Figure~\ref{fig:prompts} (in the Appendix), we used one system prompt (requesting debugging assistance but no code outputs) with two similar user prompt options. ``Error \& Code (E\&C)'' is a straightforward option which provides both the error from the log file, plus the entire faulty code file. However, given that LLMs are known to be poor at word and line counting, we also examine a second prompt, ``Error, Code, \& Line (EC\&L),'', which also reproduces the tool-localised faulty code line a second time for emphasis. 

\begin{table*}
    \centering
    \caption{Aggregated pedagogical grades for generated explanations grouped by IDEs, Language, Prompt Strategies, and LLMs}
    \vspace{-3mm}
    \label{tab:results}
    \resizebox{\linewidth}{!}{\input{fig/results}}
    \vspace{-3mm}
\end{table*}

\textbf{Response Generation:}
OpenAI's LLMs are non-deterministic, potentially giving different outputs for the same inputs. However, the GPT4 models are also more expensive to run. We decided to run the \texttt{\small gpt-3.5-turbo} model 10 times for each bug, and the other models just once, meaning that for each bug in Table~\ref{tab:bugs} we prompted for LLM responses 48 times (2 IDEs $\times$ (10 iterations of \texttt{\small gpt-3.5-turbo} and 1 each of \texttt{\small gpt-4} and \texttt{\small gpt-4-turbo-preview})). However, as noted in Table~\ref{tab:bugs}, during experimentation we found that certain bugs (5, 10, and 17) were errors in only one IDE; in total we collected 936 LLM responses for grading.

\subsection{Manual grading with pedagogically focused metrics}
\label{sec:method-metrics}

It is difficult to automatically judge the quality of answers by a question and answer system reliably. 
In this work we therefore \textit{manually} grade each of the 936 LLM-generated explanations against a series of metrics based on those used in \cite{taylor_dcc_2024}. 
To avoid complexity/marker subjectivity, we only grade using binary yes/no questions, and to avoid inter-rater reliability challenges all answers were graded uniformly by the first author. 
Our metrics follow: 
\begin{itemize}[leftmargin=*]
    \item \textit{Concept accurate}: i.e. Does the explanation link to the right concepts and keywords?
    \item \textit{No inaccuracies}: i.e. Does the explanation only contain factually correct information? An explanation may be accurate even if it is incomplete. Falsehoods can lead learners astray. %
    \item \textit{Relevant:} i.e. Is the explanation relevant to the problem at hand? 
    Whether the explanation is correct or incorrect does not impact the relevance assessment. 
    \item \textit{Correct \& complete}: i.e. Does the explanation contain everything a user needs to understand and fix the error? This is the metric we use to grade overall success.
    \item \textit{Solution is provided}: i.e. Did the model provide `too much' help? From the constructivism pedagogy, we know that learners build knowledge better by `doing' rather than by being directly `told'~\cite{ben-ari_constructivism_2001}. To judge this category, we answer `Yes' if an answer was provided with code that could be copied and pasted (even if the overall answer was wrong).
\end{itemize}

Figure~\ref{fig:explanation} presents two example explanations from \texttt{\small gpt-3.5-turbo} for Bug 1 from Figure~\ref{fig:bad-error} (a), one judged as `good' and one `bad'.
The `good' correctly identifies the error, and provides a detailed (and complete) method to fix it. However, the `bad' misidentifies the error and suggests adding a ``then" keyword incorrectly.
Figures~\ref{fig:explanation:too-much-help} and~\ref{fig:explanation:relevant-vs-concept-accurate} in the Appendix show examples of other metric scores.

\begin{figure}[h]
    \centering
    \begin{lstlisting} 
The bug is in the process statement of the VHDL code. The issue is with the missing semicolon after `data_out <= (others => '0')` within the `if rst = '1' then` block. The error message indicates that the VHDL syntax is expecting a semicolon before the `elsif` keyword. Adding the missing semicolon will resolve the syntax error.
    \end{lstlisting}
    \vspace{-2mm}
    {\footnotesize (a) `Good' explanation---Concept accurate: \textbf{Yes}; Any inaccuracies: \textbf{No}; Relevant: \textbf{Yes}; Correct \& complete: \textbf{Yes}, Solution is provided: \textbf{No}}
    \begin{lstlisting}
The bug in the provided VHDL code is that there is a missing "then" keyword after the condition `rst = '1'` in the `if` statement. This causes a syntax error when the `elsif` statement is encountered because the previous `if` statement is not properly closed.

To fix this bug, you should add the `then` keyword after the condition `rst = '1'` in the `if` statement.
    \end{lstlisting}
    \vspace{-2mm}
    {\footnotesize (b) `Bad' explanation---Concept accurate: \textbf{No}; Any inaccuracies: \textbf{Yes}; Relevant: \textbf{No}; Correct \& complete: \textbf{No}, Solution is provided: \textbf{No}}
    \vspace{-1mm}
    \caption{Example of `good' and `bad' error explanations for Bug 1 (Figure~\ref{fig:bad-error}) generated by \texttt{gpt-3.5-turbo}. Each bug is presented with graded metrics.}
    \label{fig:explanation}
    \vspace{-4mm}
\end{figure}

%% file: fig/error-list.tex
\begin{tabular}{llll|ll|ll|}
\cline{5-8}
 &
   &
   &
   &
  \multicolumn{2}{l|}{\textbf{\begin{tabular}[c]{@{}l@{}}LLM answer:\\ concept accurate\end{tabular}}} &
  \multicolumn{2}{l|}{\textbf{\begin{tabular}[c]{@{}l@{}}LLM answer:\\ correct \& complete\end{tabular}}} \\ \hline
\multicolumn{1}{|l|}{\textbf{Bug}} &
  \multicolumn{1}{l|}{\textbf{Error type}} &
  \multicolumn{1}{l|}{\textbf{Language}} &
  \textbf{Error description} &
  \multicolumn{1}{l|}{\textbf{E\&C}} &
  \textbf{EC\&L} &
  \multicolumn{1}{l|}{\textbf{E\&C}} &
  \textbf{EC\&L} \\ \hline
\multicolumn{1}{|l|}{1} &
  \multicolumn{1}{l|}{Syntax error} &
  \multicolumn{1}{l|}{VHDL} &
  Missing semicolon &
  \multicolumn{1}{l|}{71\%} &
  67\% &
  \multicolumn{1}{l|}{46\%} &
  58\% \\ \hline
\multicolumn{1}{|l|}{2} &
  \multicolumn{1}{l|}{Type error} &
  \multicolumn{1}{l|}{VHDL} &
  Can't add std\_logic\_vectors &
  \multicolumn{1}{l|}{100\%} &
  96\% &
  \multicolumn{1}{l|}{21\%} &
  25\% \\ \hline
\multicolumn{1}{|l|}{3} &
  \multicolumn{1}{l|}{Compilation error} &
  \multicolumn{1}{l|}{VHDL} &
  Can't write to an input ports object &
  \multicolumn{1}{l|}{100\%} &
  100\% &
  \multicolumn{1}{l|}{79\%} &
  83\% \\ \hline
\multicolumn{1}{|l|}{4} &
  \multicolumn{1}{l|}{Width mismatch} &
  \multicolumn{1}{l|}{VHDL} &
  Mismatch in the size of two std\_logic\_vectors &
  \multicolumn{1}{l|}{100\%} &
  100\% &
  \multicolumn{1}{l|}{71\%} &
  79\% \\ \hline
\multicolumn{1}{|l|}{5*} &
  \multicolumn{1}{l|}{Type conversion} &
  \multicolumn{1}{l|}{VHDL} &
  Can't perform two operations simultaneously in one line &
  \multicolumn{1}{l|}{100\%} &
  92\% &
  \multicolumn{1}{l|}{58\%} &
  42\% \\ \hline
\multicolumn{1}{|l|}{6} &
  \multicolumn{1}{l|}{Signal and variable} &
  \multicolumn{1}{l|}{VHDL} &
  Declaring a variable outside of a subprogram or process &
  \multicolumn{1}{l|}{100\%} &
  100\% &
  \multicolumn{1}{l|}{63\%} &
  63\% \\ \hline
\multicolumn{1}{|l|}{7} &
  \multicolumn{1}{l|}{Concurrent and sequential error} &
  \multicolumn{1}{l|}{VHDL} &
  Having both `wait' and a sensitivity list in the same process &
  \multicolumn{1}{l|}{71\%} &
  67\% &
  \multicolumn{1}{l|}{50\%} &
  38\% \\ \hline
\multicolumn{1}{|l|}{8} &
  \multicolumn{1}{l|}{Semantic error} &
  \multicolumn{1}{l|}{VHDL} &
  Using a signal or variable that has not been declared &
  \multicolumn{1}{l|}{100\%} &
  100\% &
  \multicolumn{1}{l|}{88\%} &
  88\% \\ \hline
\multicolumn{1}{|l|}{9} &
  \multicolumn{1}{l|}{Signal Readability error} &
  \multicolumn{1}{l|}{VHDL} &
  Attempting to read from an object with the mode ``out" &
  \multicolumn{1}{l|}{100\%} &
  100\% &
  \multicolumn{1}{l|}{96\%} &
  96\% \\ \hline
\multicolumn{1}{|l|}{10*} &
  \multicolumn{1}{l|}{Top Level Undefined} &
  \multicolumn{1}{l|}{VHDL} &
  Incorrect definition of the top-level module or entity &
  \multicolumn{1}{l|}{100\%} &
  100\% &
  \multicolumn{1}{l|}{83\%} &
  92\% \\ \hline
\multicolumn{1}{|l|}{11} &
  \multicolumn{1}{l|}{Case error} &
  \multicolumn{1}{l|}{VHDL} &
  Missing certain choices in a case statement &
  \multicolumn{1}{l|}{100\%} &
  100\% &
  \multicolumn{1}{l|}{83\%} &
  79\% \\ \hline
\multicolumn{1}{|l|}{12} &
  \multicolumn{1}{l|}{Singal Bit error} &
  \multicolumn{1}{l|}{VHDL} &
  Mismatch between a std\_logic type and a string literal &
  \multicolumn{1}{l|}{58\%} &
  100\% &
  \multicolumn{1}{l|}{29\%} &
  92\% \\ \hline
\multicolumn{1}{|l|}{13} &
  \multicolumn{1}{l|}{Syntax error} &
  \multicolumn{1}{l|}{Verilog} &
  Missing semicolon &
  \multicolumn{1}{l|}{100\%} &
  100\% &
  \multicolumn{1}{l|}{79\%} &
  92\% \\ \hline
\multicolumn{1}{|l|}{14} &
  \multicolumn{1}{l|}{Semantic error} &
  \multicolumn{1}{l|}{Verilog} &
  Using an undeclared variable or signal &
  \multicolumn{1}{l|}{100\%} &
  100\% &
  \multicolumn{1}{l|}{83\%} &
  92\% \\ \hline
\multicolumn{1}{|l|}{15} &
  \multicolumn{1}{l|}{Wire and Reg error} &
  \multicolumn{1}{l|}{Verilog} &
  Assign a value declared as wire using non-blocking assignments &
  \multicolumn{1}{l|}{100\%} &
  100\% &
  \multicolumn{1}{l|}{83\%} &
  92\% \\ \hline
\multicolumn{1}{|l|}{16} &
  \multicolumn{1}{l|}{Blocking and non-blocking} &
  \multicolumn{1}{l|}{Verilog} &
  Mixing blocking and non-blocking assignments to the same variable &
  \multicolumn{1}{l|}{79\%} &
  71\% &
  \multicolumn{1}{l|}{58\%} &
  67\% \\ \hline
\multicolumn{1}{|l|}{17*} &
  \multicolumn{1}{l|}{Multiple Driver error} &
  \multicolumn{1}{l|}{Verilog} &
  Assigning different values to the same signal from different processes &
  \multicolumn{1}{l|}{100\%} &
  100\% &
  \multicolumn{1}{l|}{67\%} &
  83\% \\ \hline
\multicolumn{1}{|l|}{18} &
  \multicolumn{1}{l|}{Port error} &
  \multicolumn{1}{l|}{Verilog} &
  Connect a port that does not exist &
  \multicolumn{1}{l|}{100\%} &
  100\% &
  \multicolumn{1}{l|}{63\%} &
  63\% \\ \hline
\multicolumn{1}{|l|}{19} &
  \multicolumn{1}{l|}{Binary error} &
  \multicolumn{1}{l|}{Verilog} &
  Using an illegal character in a binary number representation &
  \multicolumn{1}{l|}{100\%} &
  100\% &
  \multicolumn{1}{l|}{75\%} &
  100\% \\ \hline
\multicolumn{1}{|l|}{20} &
  \multicolumn{1}{l|}{Infinite combinational loop} &
  \multicolumn{1}{l|}{Verilog} &
  Having a infinite combinational loop that cannot be resolved &
  \multicolumn{1}{l|}{100\%} &
  100\% &
  \multicolumn{1}{l|}{58\%} &
  71\% \\ \hline
\multicolumn{1}{|l|}{21} &
  \multicolumn{1}{l|}{Double-edge error} &
  \multicolumn{1}{l|}{Verilog} &
  Mismatch between operands used in condition of an always block &
  \multicolumn{1}{l|}{100\%} &
  100\% &
  \multicolumn{1}{l|}{83\%} &
  83\% \\ \hline 
  \multicolumn{8}{l}{\textit{* Bug 5 only an error in Vivado. Bugs 10 and 17 only an error in Quartus.}} \\
\end{tabular}

%% file: fig/results.tex
\begin{tabular}{|l|c||c|c||c|c||c|c||c||c|c|c|}
\hline
\textbf{Measurement}          & \textbf{Total} & \textbf{\begin{tabular}[c]{@{}c@{}}IDE\\ Vivado\end{tabular}} & \textbf{\begin{tabular}[c]{@{}c@{}}IDE\\ Quartus\end{tabular}} & \textbf{\begin{tabular}[c]{@{}c@{}}Lang.\\ VHDL\end{tabular}} & \textbf{\begin{tabular}[c]{@{}c@{}}Lang.\\ Verilog\end{tabular}} & \textbf{\begin{tabular}[c]{@{}c@{}}Prompt\\ E\&C\end{tabular}} & \textbf{\begin{tabular}[c]{@{}c@{}}Prompt\\ EC\&L\end{tabular}} & \textbf{\begin{tabular}[c]{@{}c@{}}GPT3.5-t.\\ pass@10\end{tabular}} & \textbf{\begin{tabular}[c]{@{}c@{}}GPT3.5-t.\\ pass@1\end{tabular}} & \textbf{\begin{tabular}[c]{@{}c@{}}GPT4\\ pass@1\end{tabular}} & \textbf{\begin{tabular}[c]{@{}c@{}}GPT4-t.-p.\\ pass@1\end{tabular}} \\ \hline
\textbf{\# responses (n)}     & 936            & 456                                                           & 480                                                            & 528                                                           & 408                                                              & 468                                                            & 468                                                             & 780                                                                     & 78                                                                      & 78                                                             & 78                                                                           \\ \hline
\textbf{Concept accurate}     & 94.23\%        & 92.32\%                                                       & 96.04\%                                                        & 92.05\%                                                       & 97.06\%                                                          & 93.80\%                                                        & 94.66\%                                                         & 93.33\%                                                                 & 93.59\%                                                                 & 98.72\%                                                        & 98.72\%                                                                      \\ \hline
\textbf{No inaccuracies}      & 91.03\%        & 88.82\%                                                       & 93.12\%                                                        & 86.93\%                                                       & 96.32\%                                                          & 89.74\%                                                        & 92.31\%                                                         & 89.62\%                                                                 & 87.18\%                                                                 & 98.72\%                                                        & 97.44\%                                                                      \\ \hline
\textbf{Relevant}             & 84.18\%        & 84.87\%                                                       & 83.54\%                                                        & 85.42\%                                                       & 82.60\%                                                          & 81.62\%                                                        & 86.75\%                                                         & 88.97\%                                                                 & 83.33\%                                                                 & 60.26\%                                                        & 60.26\%                                                                      \\ \hline
\textbf{Correct \& complete}  & 71.26\%        & 69.74\%                                                       & 72.71\%                                                        & 66.48\%                                                       & 77.45\%                                                          & 67.31\%                                                        & 75.21\%                                                         & 74.36\%                                                                 & 69.23\%                                                                 & 53.85\%                                                        & 57.69\%                                                                      \\ \hline
\textbf{Solution is provided} & 3.31\%         & 3.51\%                                                        & 3.13\%                                                         & 2.27\%                                                        & 4.66\%                                                           & 4.27\%                                                         & 2.35\%                                                          & 2.82\%                                                                  & 2.56\%                                                                  & 3.85\%                                                         & 7.69\%                                                                       \\ \hline
\end{tabular}

%% file: sec/04results.tex
\section{Results}
\label{sec:results}

\subsection{Top-line results:}
Of the graded metrics, we consider the two most important categories `Concept accurate' (i.e. the LLM linked to the right error concepts) and `Correct \& complete' (i.e. the LLM provided everything the user needed to fix the problem).
Results per-bug for these are presented in Table~\ref{tab:bugs}'s last four columns. 
This shows that some bugs were easier to explain than others---e.g. Bug 19 was correctly explained in 100\% of cases. This is likely because certain bugs such as this one are conceptually simpler, and more likely to be issues in other languages as well, compared to say Bug 7, an error unique to VHDL and only explained correctly in 38\% of cases.

Table~\ref{tab:results} presents aggregated metrics across IDEs, Languages, Prompting strategies, and LLMs.
We see that `Correct \& complete' can be thought of as a subset of the other metrics, i.e. it may be possible to be relevant and accurate but still not feature a complete answer. 
Overall, conceptually accurate explanations were observed in 94\% of cases, with slight variations across different contexts. We saw only rare occasions where explanations featured outright mistakes (No inaccuracies in $\approx$91\% of cases), and overall, the LLMs had correct and complete answers in $\approx$71\% of cases. 

\textbf{IDEs:} Quartus sees better explanations than Vivado, indicating that the information provided in  Quartus's error messages may be of higher quality. When we performed an informal examination of this, we felt this to be true---for instance, Quartus's error message for Bug 1 (in Figure~\ref{fig:bad-error} (a)) includes the words `missing semicolon', unlike Vivado.

\textbf{Language differences:} Interestingly, errors in Verilog seem to be better explained than those in VHDL---we theorize this could be because of the relative differences in training data available online (Verilog is more popular for open-source).

\textbf{Prompting strategies:} When comparing the prompting strategies, we see that prompts that include the specific error line (EC\&L) tend to yield better responses ($\approx$75\%) compared to those that do not ($\approx$67\%)---i.e., providing the extra context and information to the language models appears to help. 

\textbf{LLMs:} To fairly compare the LLMs, we tabulated just the first responses (i.e. `pass@1') received by \texttt{\small gpt-3.5-turbo} alongside the responses by \texttt{\small gpt-4} and \texttt{\small gpt-4-turbo-preview}. Counter-intuitively, the smaller model (GPT-3.5) outperforms the two larger models in `correct \& complete', but the larger models are better at returning conceptually accurate responses without inaccuracies, although they have a greater tendency to over-help.

\subsection{Discussion}
LLMs generate their responses based on the data they have been trained over and on their ability to retain that data. For hardware as compared to software, there is much less training data online~\cite{thakur_benchmarking_2023}. Still, when we compare our generated HDL explanations with the C error explanations from \cite{taylor_dcc_2024}, we see that both works have $\approx$71\% correct \& complete, indicating that for this use case the data gap may not be significant.

Interestingly, our HDL explanations have a much lower incident rate of over-help (`Solution is provided') than in \cite{taylor_dcc_2024}---just 3.35\% compared with their 48\%. 
It is not immediately clear why this could be the case, as they also instructed GPT-3.5 to not emit answers directly. 
Perhaps OpenAI's LLMs `understand' C better, and so inadvertently are aligned to give out too much help. 
If model size can be thought of as a proxy for `intelligence', then this can also be observed with the larger model sizes in our work, where the GPT-4 models (which may `understand' the code better) had a higher rate of over-help compared to the smaller GPT-3.5.

%% file: sec/05conclusions.tex
\section{Conclusion}

This work set out to examine if LLMs could explain the kinds of synthesis errors that novice users of EDA tools will encounter. 
Our findings suggest that they indeed can, with 18/21 explored errors seeing good explanations in a majority of the LLM responses and 71\% of explanations being complete and correct overall. 
This work serves as a valuable proof of concept for LLM-powered techniques for improving the accessibility of EDA tools like Vivado and Quartus, and we believe that additional research in this area could significantly change how EDA tools are both learned and utilised by both novice and experienced engineers.